# Energy Saving and Survival Routing Protocol for Mobile Ad Hoc Networks

Baisakh
Department of Computer Science and Engineering,
Jaypee University of Engineering and Technology,
Guna, Madhya Pradesh, India

Nileshkumar R. Patel
Department of Computer Science and Engineering,
Jaypee University of Engineering and Technology,
Guna, Madhya Pradesh, India

## ABSTRACT

In this paper we propose a method to enhance the life time as well as improve the performance of the mobile ad hoc networks (MANET). Since MANET consists of devices that run on batteries, having limited amount of energy and due to the self-configuring and dynamic change of topology, all operations are performed by the node itself. More ever if any new technology and advancement are introduced in the MANET then the overhead of computation will also be performed by the individual nodes. All these computation will consume a lot of battery energy during the process of communication between sources to destination. In such scenario, we have considered DSR routing protocol as our base protocol and we attempt to make some modification on it which acts into an efficient energy saving and survival DSR (ESSDSR). We have considered DSR because it is one of the protocol which does not take energy into account and once the dedicated path is established between source to destination then it will keep sending through that path until the link is broken due to any of the intermediate node dies out of energy or mobility of the node away from its neighbor nodes and so it is considered as one of the unconventional routing protocol. Whereas ESSDSR acts not only as an energy efficient routing protocol but also shows an energy survival instinct. It establishes a path from source to destination where packet transmission can be sent for a longer period of time through the nodes having high level of residual battery power. It also informs the source node if any node has low battery energy, so that a new path can be discovered for the same destination prior the path get disconnected and data transmission get affected. And so the number of packet drops and retransmission can be reduced. Hence we can conclude that our proposed method enhances the performance of the networks as well as enhances the network life time. We have implemented our proposed protocol in ns-2.34 and evaluated the life time of the networks as well as the node's life time has been improved as compare to traditional DSR with a higher ratio.

## General Terms

Mobile Ad Hoc Networks

## Keywords

MANET, Energy Consumption in MANET, DSR, ESSDSR.

## 1. INTRODUCTION

Ad Hoc Network is a multi-hop wireless networks which consists of autonomous mobile nodes interconnected by means of wireless medium without having any fixed infrastructure. It's quick and easy deployment in a situation where it's highly impossible to set up any fixed infrastructure networks, has increased the potential used in different applications in different critical scenarios. Such as battle fields, emergency disaster relief, conference and etc. A mobile ad hoc network [MANET] can be characterized by the mobile nodes which have freedom to move at any direction and have the ability of self-configuring, self-maintaining and self-organizing themselves within the network by means of radio links without any fixed infrastructure like based station, fixed link, routers, and centralized servers. And so the whole functionality along with routing mechanisms are incorporated in every node which indeed consume a lot of battery power. Other process like topological updation when a node moves out of the network, sending and receiving packets, processing packets and routing the packet through their neighboring node also make use of heavy power consumption [1, 2]. So we assume in MANET the power gets consumed in mainly two ways. First, due to transmitting data to a desired recipient. Secondly, mobile node might offer itself as an intermediate forwarding node in the networks. The power level of the node are also getting affected the ease with which route can be established between two end points.

## 1.1 Energy Consumption in MANET

The routing protocols play a significant role in mobile ad hoc networks as the nodes are dynamic in nature and each node can participate in routing the data packets. In such scenario, energy efficient routing protocols are needed for Ad Hoc networks, especially when there are no routers, no base stations and no fixed infrastructure [3, 4]. So to establish the correct and efficient routes from a source to destination is not the ultimate aim of any routing protocols, rather to keep the networks functioning as long as possible without any interruption and with less battery consumption at each node level, should also be the objective for these routing protocols.

These goals can be accomplished by minimizing mobile node's energy during both the active as well as inactive communications. Active communication is when all the nodes of the route are participating in receiving and forwarding of data. Minimizing the energy during active communication is possible through two different approaches.

- Transmission power Control
- Load distribution

In inactive communication the nodes are idle i.e. neither forwarding any data packets nor receiving any data packets. In such situation, to minimize the energy consumption Sleep/Power down approach is used. We will not discuss about the power consumption during inactive communication in the network. There are many energy matrices used for calculating the power consumption caused by different reasons. The energy few energy related metrics are used. These metrics are helpful while determining energy efficient routing path instead of considering shortest path like in the traditional DSR protocol use.

These metrics are:

- Energy consumed per packet

- Time to network partition

Variation in node power level is:

- Cost per packet
- Maximum node cost

The tradeoff between frequency of route update dissemination and battery power utilization is one of the major design issues of ad hoc network protocols. In ad hoc networks, mainly the routing protocols are classified into two types on the basis of the routing information update mechanism the protocols are divided into two types:

- Table driven (proactive) Protocol
- On-demand (reactive) Protocol

In the reactive protocol every individual node maintains the route for the entire networks whereas in reactive protocol, a route to a certain destination is established when there is a demand from a source for that particular destination. And so the source initiates it by invoking the Route Discovery Method. Dynamic Source Routing (DSR) protocol belongs to the reactive protocol class.

As the MANET nodes are operated by limited battery power, this power should be effectively used so that node's life span can be maximized. But DSR is a kind of protocol which does not concern about the power consumption of battery and lacks the power aware mechanism. So here in this paper we have made an attempt to convert DSR from an unconventional protocol to an energy efficient routing protocol by making some modification. We have proposed our protocol as ESSDSR (energy saving and survival DSR).

The remainder of this paper is structured as follows. Since ESSDSR is based on DSR, a brief description of DSR is overviewed in the section II. The section III presents an energy saving and survival DSR (ESSDSR). Section IV will present the proposed algorithm ESSDSR. And the last section IV will contain a summary of our proposed method and future works.

## 2. DYNAMIC SOURCE ROUTING (DSR)

### 2.1 DSR Routing Mechanism

Dynamic Source Routing (DSR) [1, 2, 13, 14] is a simple and efficient routing protocol designed specification for use in multi-hop wireless ad hoc mobile networks. DSR is one of the important routing protocols that are used for mobile ad hoc networks as many energy efficient routing protocols [3, 4, 5, 5, 7, 8, 9, 10] are designed based on its mechanism. It finds the route from source to destination only when the source initiates route discovery process. All aspects of protocol operate entirely on demand. This protocol also makes the network self-organizing and self-configuring. Basically the protocol is composed of two mechanisms, Rote Discover and Route Maintenance and these two mechanisms work together to allow nodes to discover and maintain the source route to any destination node in the a hoc networks.

### 2.2 Route Discovery

Route discovery is done with two sub steps that is, Route Request (RREQ) and Route Reply (RREP).

#### 2.2.1 Route Request:

The route discovery comes in play when a mobile node has some data/packet to send to any destination and it does not have any route to the destination in its route cache. Then it initiates route discovery by broadcasting a route request packet. This route request contains address of the destination, address of the source and a unique identification number that is generated by the source node only. Each node receives the packet and checks whether the packet is meant for it or not. If it is not the destination node then it simply forwards the packet to the outgoing links adding its own address in the packet. To avoid duplicate route request which is generated from the same source, a node only forwards the route request that has not yet been seen appear in the route request with the same identification number.

#### 2.2.2 Route Reply

As soon as the packet arrives at the destination node or arrives at a node that contains in its route cache an unexpired route to the destination, then a route reply is generated. Not only the packet contains all the adder of the intermediate node it has come across but the sequence of hops is also stored in it. The Route reply is generated by the destination placing the route record contained in the route request into route reply. During the route reply if the destination node has the route to the initiator in its route cache, it may use that route for route reply. Otherwise destination node may reverse the route in the route record if the link is symmetric. If the symmetric links are not supported then the node may initiate its own route discovery piggybacking the route reply on the new route request. When any intermediate node receives any route reply from destination node or any other node then they append their route record and forward it to its neighbor nodes.

#### 2.2.3 Route Maintenance

Route maintenance is a process of identifying link whether it is reliable and capable of carrying packet on it or not. This process is executed by the use of route error packets and acknowledgements. When the data link layer encounters a fatal transmission problem then a route error message is generated. Suppose a packet is retransmitted (up to a maximum number of attempts) by some hop the maximum number of times and number of receipt conformation in received, then this node returns a packet error message to the original sender of the packet, identifying the link over which the packet could not be forwarded.

#### 2.2.4 Benefits and Limitations

As the entire route is contained in the packet header, there is no need of having routing table to keep route for a given packets. The caching of any initiated or overheard routing data can significantly reduce the number of control message being sent, reducing overhead.

But DSR requires significantly more processing resources than most of other protocols. The other drawback of DSR is selecting the path for routing on the basis of minimum hop counts from the source to the destination. As it selects the path of having minimum hops count, lesser will be the number of intermediate nodes, more will be the distance between each pair of nodes. As the distance is more we need to have more transmission power to communicate between any pair of nodes and hence it consumes more battery power as it is one of the limited resources.

# 3. ENERGY SAVING and SURVIVAL DSR (ESSDSR)

## 3.1 ESSDSR Energy Saving Mechanism

The objective of ESSDSR is to forward the packet through those nodes which are having higher level of energy at a given time. The DSR has been modified in such a way that if an energy of a node which is forwarding the data packet within a multi hop path reaches a level less than or equal to certain threshold percentage of its initial battery energy the node will ask the neighbor nodes to look for another location for such data. Since the node may die out of energy in sort of time if it continues sending or receiving packets. So the modified algorithm has also introduced not only an energy saving feature but also introduces an energy survival characteristic for any low energy node.

In the traditional DSR during route Discovery process if any node receives a Route Request (RREQ) which is not meant for it or if the node is not the final destination then it keeps the packet for a certain time interval which is selected pseudo-randomly from a uniform distribution of probability between 0 to 0.01. Usually 0.01 is considered as constant broadcast jitter.

The idea behind the ESSDSR is to introduce the delay dynamically for this control packet like RREQ and RREP (Route reply) on the basis of remaining residual battery power of the intermediate nodes. The more will be residual power the less will be the delay. If any node is having low energy than certain value, then the delay in our proposed method is inversely proportional to the residual battery power. Because, traditional DSR uses the pseudo-random delay and maximum of 0.01 sec. In ESSDSR if the energy of a node is higher than the minimum level of energy ( low energy), then the delay should be less than 0.01 sec. Otherwise the delay will be maximum which is 0.01 sec. the minimum value of the energy is set to 1 joule and it corresponds to the delay of $1/(1*100)= 0.01$ sec. In our simulation we have set the initial energy of the node as 10juole. So the delay introduced at RREQ packet is $1/(10*100)=0.001$ sec.

Hence the best route for any packet will be maximum sum of total energy of the intermediate nodes among all possible paths from the same source to destination. And as the delay is inversely proportional to the residual energy, the RREQ will be passed through smallest delay and so RREQ will reach earlier to the destination than other possible path from the same source to destination. Because the length of the delay packet affects forwarding, so it ensues the longer path but with higher energy so that the data transmission can be carried out for higher period of time without any interruption unless the intermediate nodes moves out from its original position. But we have assumed our network to be a static one.

## 3.2 ESSDSR Energy Survival Mechanism

As in traditional DSR there is no concept of power consumption node and as we have proposed our method on the basis of energy consumption, we need to maintain a balance in energy consumption of intermediate nodes and consequent loss links.

Our method has provided a solution for this problem. When a node is having remaining residual energy that is  less than or equal  to certain threshold percentage of its initial energy and it has some data to send or forward then the sending node will broadcast a control packet where the header contains a flag " low energy" set to 1. In this way it informs to its all neighbor nodes not to send any packet through this node and remove the path containing this affected node from it route cache. The moment the source will get this packet, now it will start discovering a new path that does not contain the affected node as its intermediate node.

If any node B  reaches the energy level below the given threshold and there are packets to be sent from the node A to B, then the node B  will broadcast the  to all its neighbor node informing not to continue forwarding packet to him if there are other routes to the destination node. Each neighbor node of node "A" after getting the broadcast packet with low energy set to 1 will then remove the path from A to B from its route cache.

## 3.3 ESSDSR Algorithm

Step-1 At the beginning of the communication from source to destination, the route discovery will be done according     to the traditional DSR  routing protocol where the dedicated path will be chosen on the basis of the minimum hop count as all the nodes are having same initial energy, assumes to be 100% of battery power.

Step-2 Whenever the energy of any node reaches an energy level less than or equal to the certain threshold then low energy field in the DSR header packet will be set to 1 and broadcast to its neighbor nodes.

Step-3 When the neighbor node of the affected node receives the broadcast packet sent by the affected node,     then they will remove the path to that node from its route cache and broadcast an error (Route Error) to the source.

Step-4 The moment source will receive the error message it will start route discovery process finding out the path from source to destination without containing the affected node.

Step-5 Now the delay will be introduced to RREQ packet according to the remaining energy power of the battery. So the node with higher energy will have lesser delay and reaches out early to the destination.

# 4. EXPERIMENTAL RESULTS

We have evaluated the individual node's life time and network life time in our experiment which is being done by ns- 2.34 [15, 16] to perform our simulation and compared the traditional and our proposed ESSDSR. In two cases our method has shown better results than the DSR: one in enhancing the life time of the individual node and the total network life time.

The simulation network consists of 11 nodes which are static with the range of 300x200 m2 area. We have used here the energy model as well as the propagation model for calculating the node energy and dynamic transmission power of individual node. FTP traffic is introduced between the node 0 which is the source node and node 11 which is the destination node and they are not connected directly. The initial energy of the node was kept as 10 Joules and we have varied the initial energy from 10-20 Joules across the whole 11 nodes. We have set all the even node number with 20 Joules and odd node number with 10 Joules of initial energy.

## 4.1 Individual node's lifetime

We have simulated both the DSR and ESSDSR and  at the end of the simulation, the remaining energy of the individual node is shown in the figure-1. The node 1, 3, 5, and 7 are completely used and their energy level became zero. The rest of the nodes which were partially used are having higher energy level than that of original DSR. The life time of these nodes are drastically improved and as their remaining energy is more in ESSDSR as compare to DSR, they can be used further for the data communication.

| Table Head | Simulation Parameter | Value |
|---|---|---|
| 1 | Network Size | 300x200 |
| 2 | Number of Nodes | 11 |
| 3 | Simulation Duration | 60sec |
| 4 | Traffic type | FTP type |
| 5 | Packet Size | 1080 (TCP) 40 (ACK) in bytes |
| 6 | Queue type | Drop Tail |
| 7 | Propagation Model | Two Ray propagation Model |
| 8 | Antenna | Unidirectional |
| 9 | Transmission Range | 250m |
| 10 | rxPower | 0.925W |
| 11 | txPower | 1.43W |
| 12 | sleepPower | 0.045W |

## 4.2 Network Lifetime

From the figure-2, It is shown that the network life time has been more doubled. From the experiment, we calculate the network life time of the network by ESSDSR is 49.831 second while the network life time by DSR is 31. 016 sec. So there is 61.71 percentage of improvement in total life span of the network. This is due to the energy saving and survival mechanism of ESSDSR where the route is chosen on the basis of higher residual battery power of the intermediate nodes instead of the nodes having less energy. Because such nodes may cause more number of network partition and more number of packet retransmission. Hence due to broken path, a new route is required which can be obtained by route discovery procedure and this is consumes more battery power.

## 5. CONCLUSION and FUTURE WORK

In this paper, we have designed an energy saving and survival DSR routing protocol which is based on DSR, where we have incorporated the energy saving technique in every node and introduced a method like energy survival technique which alerts its source and neighbor node about its low energy by broadcasting the special packet carrying Low Energy which then set as 1.

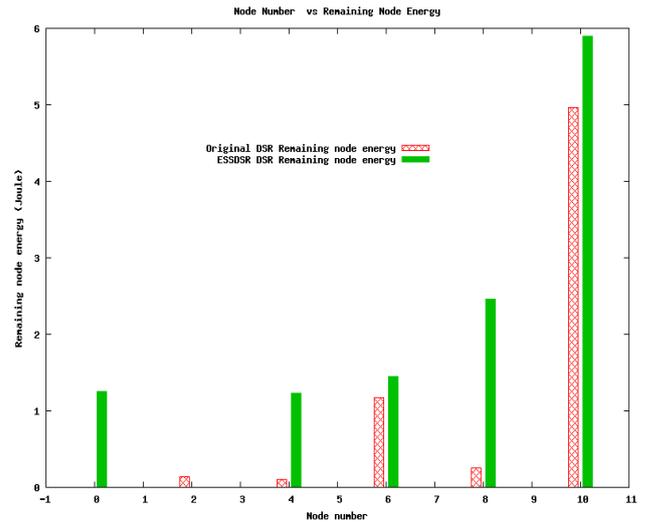

**Figure-1: Energy consumption of nodes**

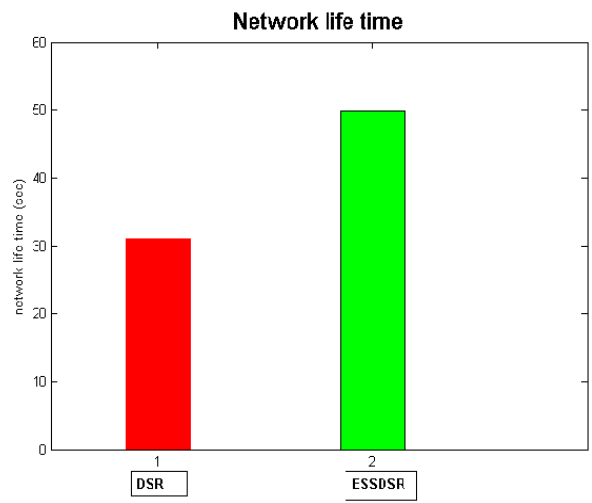

**Figure-2 Network lifetime**

In this way it avoids such low energy nodes from overusing which may lead them to die soon. Our method has improved the individual node life time and so the whole network life span. And it has chosen its first path with the higher traffic time as the selection procedure is based on the residual power of the individual nodes.

Still a lot of analysis is to be done to evaluate the performance of ESSDSR with respect to the original DSR as we have not considered some of the parameters like network throughput, end to end delay, packet loss, packet delivery ratio etc. We can also carry out the same method with more number of nodes with multiple sources and destinations at the same time during the simulation. Mobility factor may be introduced in the nodes and further experiments can be done as future works.

## 6. REFERENCES

[1] C. Siva Ram Murthy, B. S. Manoj, "Ad Hoc Wireless Networks Architecture and Protocols", 2nd ed, Pearson Education, 2005.

[2] E. Royer and C.-k. Toh, "A Review of Current Routing Protocols for Ad Hoc Moblie Wireless Networks," IEEE Personal Comm.Magazine, vol. 6, no. 2, Apr. 1999.

[3] C. Yu, B. Lee, H. Youn, "Energy Efficient Routing Protocols for Mobile Ad Hoc Networks", Wireless Communication and Mobile Computing, Wireless Com. Mob. Computing (2003).

[4] Chang J-H, tassiulus L., "Energy Conserving Routing in Wireless Ad Hoc networks", In Proceedings of IEEE INFOCOM, March, 2000,pages 22-31

[5] Stojmenovic I., Lin X., "Power Aware Localized Routing in Wireless Networks", IEEE transaction on parallel and Distributed Systems Vol. 12 issue.11, November, 2001, pages 1122-1133.

[6] Doshi S., Bhandare s., Brown T.X., "An On-demand Minimum Energy Routing Protocol for Wireless Ad Hoc Networks", ACM SIGMOBILE Mobile Computing and Communication Review, vol. 6, Issue 2, pages 50-66.

[7] Banerjee S, Mishra A. "Minimum Energy Path for Reliable Communication in Multi-hop Wireless Network", In Proceedings of the 3d ACM Annual Workshop on mobile Ad Hoc networking and Computing (MobiHoc), June 2002, pages 146-156.

[8] Toh C.-K. "Maximum Battery Life Routing to Support Ubiqutious Mobile Computing in Wireless Ad Hoc Networks", IEEE Communication Magazine, Vol. 39, issue. 6, June 2001, pages 138-147.

[9] Xu Y., Heidemann J., Estrin D. "Geography-informed Energy Consevation for Ad Hoc Routing", in Proceedings of 7th Annual international Conference on Mobile Computing and Networking (Mobicom), july 2001, pages 70-84.

[10] Feeney, L. M., Nilsson, M., "Investigating the Energy Consumption of Wireless Network Interface in an Ad Hoc Networking Environment", IEEE INFOCOM, vol. 3, April 2001, pages 1548-1557.

[11] A. Zhou, H. hassanein, "Load-Balanced Wireless Ad Hoc Routing", Proceedings of Cnadian Conference on Electriccal and Computer Engineering, vol. 2, 1157-1161 (2001).

[12] G. Chakrabarti, S. Kulkarni, "Load Balancing and Resource Reservation in Mobile Ad Hoc Networks", Ad Hoc Networks, Volume 4, issue 2, 1 March (2006).

[13] IRTEF Draft, The Dynamic Source Routing Protocol (DSR) for Mobile Ad Hoc Networks, available at: http://tools.ietf.org/html/rfc4728 (2010).

[14] D.B johnson, D.A Maltz, Y.-C. Hu, The Dynamic Source Routing Protocol for Mobile Ad Hoc Networks (DSR). Available from: http://www.ietf.org/internet-drafts/draftietf-manet-dsr-10.txt (2007).

[15] The network simulator (NS-), http://www.isi.edu/nsnam/ns (2010).

[16] Fall. K. and varadhan, K., "NS Notes and Documentation Technical report" Ubniversity of California- berkly, LBL, USC/ISI and Xerox PARC.